\documentstyle[12pt]{article}

\newcommand{\newc}{\newcommand}
\newc{\ra}{\rightarrow}
\newc{\lra}{\leftrightarrow}
\newc{\beq}{\begin{equation}}
\newc{\eeq}{\end{equation}}
\newc{\barr}{\begin{eqnarray}}
\newc{\earr}{\end{eqnarray}}

 1

\begin{document}
\begin{titlepage}
\begin{flushright}
{IOA.312/95}\\
\end{flushright}
\begin{center}
SEARCHING FOR COLD DARK MATTER\\ \vspace{15mm}
J. D. Vergados\\ \vspace{20mm}
Theoretical Physics Section, University of Ioannina,\\
 GR 451 10, Ioannina, Greece \\

\end{center}
\vspace{10mm}
\begin{abstract}
\vspace{8pt}

The differential cross-section for the elastic scattering of the lightest
supersymmetric particle (LSP) with nuclear targets is calculated in the
context of currently fashionable supersymmetric theories (SUSY).  An effective
four fermion interaction is constructed by considering i) $Z^0$ exchange
ii)s-quark exchange and iii) Higgs exchange.  It is expressed in terms of the
form factors $f^0_V,f^0_A, f^0_S$  (isoscalar)  and $f^1_V,f^1_A$ and  $f^1_S$
(isovector)  which contain all the information of the underlining theory.
Numerical values were obtained using representative input parameters in the
constrained parameter space  of SUSY phenomenology.  Both the coherent and for
odd-A nuclei the incoherent (spin) nuclear matrix elements were evaluated for
nuclei of experimental interest.  The spin matrix elements tend to dominate for
 odd nuclei but the coherent matrix elements become more important in all
other cases.  For the coherent part the Higgs contribution competes with the Z-
and s-quark contributions.  Cross-sections as high as $10^{-37} cm^2$ have been
obtained.

 Latex\\ e-mail: vergados@cc.uoi.gr
\end{abstract}
\end{titlepage}

\vfill
\eject

\newpage

{ \it {\bf 1. Introduction}}.\\

In recent years the phenomenological implications
of supersymmetry (SUSY) are
being taken seriously [1-3].  Pretty accurate predictions  at low energies
are now feasible in terms of few input parameters in the context of unified
minimal SUSY without commitment to specific gauge groups[4-8]. More or less
such predictions do not seem to depend on arbitrary choices of some parameter
or untested assumptions [1-3,9-40]. One may not have to wait, however, till
supersymmetry is discovered in high energy colliders. Instead one may look
now  at phenomena which supersymmetry  might affect, e.g. proton decay,
lepton flavor violation [10] ($\mu \ra e \gamma$ etc) and dark matter.  In the
present paper we will concentrate on the implications  of supersymmetry on
dark matter [11-16].

There is ample evidence [17-28] that more than 90\% of the mass of our galaxy,
or even of
the whole universe, is made up of matter of unknown nature.  If one goes beyond
the
standard model of weak and electromagnetic interactions one has a number
of choices for
the dark matter candidates.  The most obvious choice is, of course,
 particles which
exist, like neutrinos if they have a mass of $\sim 10 eV$.
Such light particles are
expected to be relativistic and constitute the  Hot Dark Matter Component
(HDM).
Another possibility are the axions,  which were introduced to account for
the strong CP
problem.  Even though they have not been found in any of the experimental
searches, this
does not mean that they should be ruled out entirely.
 The third  and most appealing
possibility, linked  closely with supersymmetry, is the lightest supersymmetric
particle (LSP) which is expected to be neutral (see section 2).
This particle, which will be denoted by $\chi_1$, is expected to be massive
[1-3]
(10-100GeV) moving with non relativistic velocities (kinetic energy 10-100
KeV).
In the
absence of R-parity violating interactions such a particle is absolutely
stable. It
constitutes the Cold Dark  Matter (CDM) component.  The CDM component is needed
for
the large scale structure formation in the universe.  As a matter of fact the
ratio of
CDM to HDM [23] is 2:1.  Since one expects a baryonic component of about 10\%,
one
is lead  to the scenario
\beq
\Omega_{CDM} = 0.6, \,\,\,\Omega_{HDM} = 0.3, \,\,\, \Omega_{B} = 0.1
\label{eq:eg1}
\eeq
where
\beq
\Omega_i = \frac {\rho_i} {\rho_c}, \,\,\,\,\rho_c = \frac {3H^2} {8\pi G_N}
 \simeq 10
\frac {nucleons} {m^3} h^2_0  \label{eq:eg2}
\eeq
where $H$ is the  Hubble constant, $G_N$ is Newton's gravitational  constant
and $h_0$ lies  between 0.5 and 1.

The detection of LSP can in principle be achieved by devices which are able
to detect particles which are interacting very weakly [24-25].
This can be achieved by
detecting the recoiling nucleus in the reaction
 \beq
\chi_1 + (A,Z) \ra \chi_1 + (A,Z)  \label{eq:eg3}
\eeq
The recoiling energy can be  converted into phonon
 energy and detected as temperature
rise.  This requires a crystal at low temperatures with sufficiently high Debye
temperature.  The detector should be small enough to permit anticoincidence
shielding to reduce background and large enough to allow a sufficient number
of counts. A compromise of about 1kg is achieved.
 Another possibility is to use
superconducting granules suspended in a magnetic field.
The produced heat will
destroy the superconductor and  the resulting magnetic flux will trigger a
signal in the pick -up coil.  Again a mass of about 1 kg is optimum.

The background  of such detectors is composed of cosmic rays and natural
radioactivity. It can be tackled by utilising the so-called modulation effect,
caused by the change in the relative velocity of LSP and the detector
due to the
diurnal [26] and annual [27] motion of the Earth.

The indirect detection is another possibility.  The
LPS's trapped in the gravitational field of the sun will
 pair anihilate producing high
energy particles. Of particular interest are high energy
neutrinos originating from the
sun, which can be detected by the neutrino telescopes.

In the present paper we will calculate the cross section
for the scattering of the LSP
by a nucleus.  We will utilize the recent developments
 in supersymmetric theories
which have yielded a substantially constrained parameter space.
 In the first
step, along  the lines of ref [11-16],  we will construct from the elementary
couplings  the effective four fermion interaction which couples the LSP to the
quarks. The next step consists in writing
the relevant four fermion interaction
at the quark level. The final step consists in transforming this interaction at
the
nucleon level. In the present work we will consider Z, s-quark and Higgs
exchange. For the first two exchanges the basic
interaction can be read off from
the appendix of ref. [5]. The transformation to the nucleon level is
straightforward [28].
 One can thus construct the 4 needed form factors, i.e.
the vector and axial vector isoscalar ($f_V^0,f_A^0$)
and isovector ($f_V^1, f_A^1$)
form factors.
 Since, however, the Higgs exchange contribution is
important due to the coherent effect of all nucleons we will also provide the
model dependent scalar form factors  $f_S^0$
(isoscalar) and  $f_S^1$ (isovector).
 The  diagrams which involve Higgs exchange are a bit more model dependent
[5]. Also in this case the transition to the nucleon level is not
so straightforward [29]. We will see that the spin dependent
nuclear matrix elements arising from the Axial currents
 are important especially
for light nuclei. We will estimate them in this work and
provide more accurate
calculations in a future publication.\\ \vspace{0.2cm}

{ \it {\bf 2. Effective Lagrangian}}.\\

Before proceeding with the
construction of the effective Lagrangian we will briefly discuss
the nature of the
lightest supersymmetric particle (LSP) focusing on those ingredients
which are of interest to dark matter.\\
\vspace{0.2cm}

{ \it {\bf 2.1. The nature of LSP}} \\

In currently favorable supergravity models the LSP is a linear combination
[1-3,5]
 of the neutral four fermions ${\tilde B}, {\tilde W}_3, {\tilde H}_1$
and ${\tilde H}_2$
 which are the
supersymmetric partners of the gauge Bosons $B_\mu$ and $W^3_\mu$ and the
 Higgs scalars
$H_1$ and $H_2$.  Admixtures of s-neutrinos are expected to be negligible.

In the above basis the mass-matrix takes the form [1]
\beq
  \left(\begin{array}{cccc}M_1 & 0 & -m_z c_\beta s_w & m_z s_\beta s_w \\
 0 & M_2 & m_z c_\beta c_w & -m_z s_\beta c_w \\
-m_z c_\beta s_w & m_z c_\beta c_w & 0 & -\mu \\
m_z s_\beta s_w & -m_z c_\beta c_w & -\mu & 0 \end{array}\right)
\label{eq:eg 4}
\eeq

In the above expressions $c_W = cos \theta_W$,
$s_W = sin  \theta_W$,  $c_\beta = cos \beta$,$s_\beta = sin \beta$
   where  $tan \beta = <\upsilon_2>/<\upsilon_1>$ is the
ratio of the vacuum  expectation values of the Higgs scalars $H_2$ and
$H_1$. $\mu$ is a dimentionful coupling constant which is not specified by the
theory (not even its sign).  The parameters $tan \beta , M_1, M_2, \mu$ are
determined by a phenomenological fit to the data.  In our numerical treatment
we used the three possibilities of table VII of ref.1 which we
considered to be representative samples in the allowed by the data SUSY
parameter space (see table I).

By diagonalizing the above matrix we obtain a set of eigenvalues $m_j$ and
the diagonalizing matrix $C_{ij}$ as follows
\beq
\left(\begin{array}{c}{\tilde B}_R\\
{\tilde W}_{3R}\\
{\tilde H}_{1R} \\
 {\tilde H}_{2R}\end{array}\right) = (C_{ij})
\left(\begin{array}{c} \chi_{1R} \\
\chi_{2R} \\   \chi_{3R} \\
\chi_{4R} \end{array}\right)  ;
\left(\begin{array}{c} {\tilde B}_L \\
{\tilde W}_{2L} \\   {\tilde H}_{1L} \\
{\tilde H}_{2L} \end{array}\right) =  \left(C^*_{ij} \right)
\left(\begin{array}{c} \chi_{1L} \\
\chi_{2L} \\   \chi_{3L} \\
\chi_{4L} \end{array}\right)
\label{eq:eg.8}
\eeq
Another possibility to express the above results in photino-zino
basis ${\tilde \gamma}, {\tilde Z}$ via
\barr
{\tilde W}_3 &=& sin \theta_W {\tilde \gamma}
 -cos \theta_W {\tilde Z} \nonumber \\
{\tilde B}_0 &=& cos \theta_W {\tilde \gamma}
 +sin \theta_W {\tilde Z} \label{eq:eg9}
\earr
In the absence of supersymmetry breaking $(M_1=M_2=M$ and $\mu=0)$ the
photino is one of the eigenstates  with mass $M$.  One of the remaining
eigenstates has a zero eigenvalue and is a linear combination of ${\tilde
H}_1$ and  ${\tilde H}_2$ with mixing angle $sin \beta$. In the  presence
of SUSY breaking terms the ${\tilde B}, {\tilde W}_3$ basis is superior
since the lowest eigenstate $\chi_1$ or LSM is primarily ${\tilde B}$.  From
our point of view the most important parameters are the mass $m_1$ of LSP
and the mixings $C_{j1}, j=1,2,3,4$ which yield the $\chi_1$ content of the
initial basis states.  These parameters are given in table II.

We are now in a position to find the interaction of $\chi_1$ with matter.
We distinguish  three possibilities involving Z-exchange, s-quark exchange and
Higgs exchange.\\
\vspace{0.3cm}

2.2.1. The Z-exchange contribution \\

This
can arize from the interaction of Higgsinos with $Z$ which can be read from
eq.C86 of ref.[5]. \beq
{\it L} = \frac {g}{cos \theta_W} \frac {1}{4} [{\tilde H}_{1R}
\gamma_{\mu}{\tilde H}_{1R} -{\tilde H}_{1L}\gamma_{\mu} {\tilde H}_{1L} -
({\tilde H}_{2R}\gamma_{\mu}{\tilde H}_{2R}  -{\tilde H}_{2L}
\gamma_{\mu}{\tilde
H}_{2L})]Z^{\mu}
 \label{eq:eg 10}
\eeq
Using eq. (5) and the fact that for Majorana particles
${\bar \chi} \gamma_{\mu} \chi = 0$,  we
obtain
\beq
{\it L} = \frac {g}{cos \theta_W} \frac {1}{4} (|C_{31}|^2
-|C_{41}|^2)  ({\bar \chi}_1\gamma_{\mu} \gamma_5 \chi_1) Z^{\mu}
 \label{eq:eg 11}
\eeq
which leads to the effective 4-fermion interaction (see fig. 1a)
\beq
{\it L}_{eff} = \frac {g}{cos \theta_W} \frac {1}{4} 2(|C_{31}|^2
-|C_{41}|^2)  (-) \frac {g}{2cos \theta_W} \frac {1}{q^2 -m^2_Z}
 {\bar \chi}_1\gamma^{\mu} \gamma_5 \chi_1)J^Z_\mu
 \label{eq:eg 12}
\eeq
where the  extra factor [5] of 2 comes from the Majorana nature  of
 $\chi_1$. The
neutral  hadronic current $J^Z_\lambda$ is given by
\beq
J^Z_{\lambda} = - {\bar q} \gamma_{\lambda} ( \frac {1}{3} sin^2 \theta_W -
(\frac {1}{2} (1-\gamma_5) - sin^2 \theta_W))\tau_3 q)
  \label{eq:eg 13}
\eeq
at the nucleon level it can be written as
\beq
{\tilde J}_{\lambda}^Z = -{\bar N} \gamma_{\lambda} (sin^2 \theta_W
-g_V (\frac {1} {2} - sin^2 \theta_W) \tau_3 + \frac {1} {2}
 g_A \gamma_5 \tau_3) N
  \label{eq:eg 14}
\eeq
Thus we can write
\beq
{\it L}_{eff} = - \frac {G_F}{\sqrt 2} ({\bar \chi}_1 \gamma^{\lambda}
\gamma^5 \chi_1) J_{\lambda}(Z)
 \label{eq:eg 15}
\eeq
where
\beq
J_{\lambda}(Z) = {\bar N} \gamma_{\lambda} [f^0_V(Z) + f^1_V(Z) \tau_3
+  f^0_A(Z) \gamma_5 + f^1_A(Z) \gamma_5  \tau_3] N
\label{eq:eg 16}
\eeq
and
\barr
f^0_V(Z) &=& 2(|C_{31}|^2 -|C_{41}|^2) \frac {m^2_Z}{m^2_Z - q^2} sin^2
\theta_W \\
f^1_V(Z) &=& - 2(|C_{31}|^2 -|C_{41}|^2) \frac
{m^2_\mu}{m^2_\mu - q^2}g_V (\frac {1}{2} - sin^2 \theta_W) \\
f^0_A (Z) &=& 0  \\
f^1_A (Z) &=&  2(|C_{31}|^2 -|C_{41}|^2) \frac {m^2_Z}{m^2_Z - q^2} \,\,
\frac
{1}{2} g_A \label{eq:eg 13d}
\earr
with $g_V=1.0,  g_A = 1.24$. We can easily see that
\beq
f^1_{V}(Z)/ f^0_{V}(Z) = -g_V ( \frac {1}{2sin^2 \theta_W} - 1 ) \simeq
- 1.15  \nonumber
\eeq
Note that the suppression of this Z-exchange interaction compared to
the ordinary
neutral current interactions arises from the smallness of the mixings
$C_{31}$ and $C_{41}$, a consequence of the fact that the Higgsinos are
normally quite a bit heavier than the gauginos.  Furthermore, the two
Higgsinos tend to cancel each other.
\vspace{0.2cm}

2.2.2 The $s$-quark mediated interaction

The other interesting possibility arises from the other two components of
$\chi_1$, namely ${\tilde B}$ and ${\tilde W}_3$. Their corresponding
couplings to $s$-quarks can be read from the appendix C4 of  ref.[5].
They are
\barr
{\it L}_{eff} &=& -g \sqrt {2} \{{\bar q}_L [T_3 {\tilde W}_{3R}
- tan \theta_W (T_3 -Q) {\tilde B}_R ] {\tilde q}_L \nonumber \\
&-& tan  \theta_W {\bar q}_R Q {\tilde B}_L {\tilde q}_R\} + HC
 \label{eq:eg 17}
\earr
where ${\tilde q}$ are the scalar quarks (SUSY partners of quarks).  A
summation  over all quark flavors is understood. Of interest to us
are, of course, the flavors $u$ and $d$.
 The above interaction is to high accuracy
diagonal in quark flavor.  Using eq. (15) we can write the above equation
in the $\chi_i$  basis. Of interest to us here is the part
\barr
{\it L}_{eff} &=& g \sqrt {2} \{(tan  \theta_W (T_3 -Q) C_{11} - C_{21})
{\bar q}_L \chi_{1R} {\tilde q}_L \nonumber \\
 &+& tan \theta_W C_{11} {\bar q}_R Q \chi_{1L} {\tilde q}_R
 \label{eq:eg 18}
\earr
where Q is the charge and $T_3$ the third component of the isospin operator.
The effective four fermion interaction (fig. 1b) takes the form
\barr
{\it L}_{eff} &=& (g \sqrt {2})^2  2\{ \frac {[C_{11} tan \theta_W (T_3 -Q)
-C_{21} T_3]^2} {q^2 - m^2_{{\tilde q}_L}} ({\bar q}_L \chi_{1R})
({\bar \chi}_{1R} q_L)
\nonumber \\
 &+& \frac {|tan \theta_W C_{11} Q|^2} {q^2 - m^2_{{\tilde q}_R}} ({\bar
q}_R \chi_{1L})  ({\bar \chi}_{1L} q_R)
 \label{eq:eg 19}
\earr
The factor of two arises from the Majorana nature of $\chi_1$.
 Employing a Fierz
transformation [30],  it can be cast in a more convenient form
\barr
{\it L}_{eff} &=& (g \sqrt {2}) 2(-) \frac {1} {2} \{ \frac {|C_{11} tan
 \theta (T_3 -Q) -C_{21}T_3|^2} {q^2 - m^2_{{\tilde q}_L}} ({\bar q}_L
\gamma_\lambda q_L) ({\bar \chi}_{1R} \gamma^\lambda \chi_{1R})
\nonumber \\
 &+& \frac {|tan \theta_W C_{11} Q|^2} {q^2 - m^2_{{\tilde q}_R}}
({\bar q}_R \gamma_\lambda q_R) ({\bar \chi}_{1L} \gamma^\lambda \chi_{1L})\}
 \label{eq:eg 20}
\earr
This can be  written compactly as
\barr
{\it L}_{eff} &=& \frac {G_F} {\sqrt {2}} 2\{ {\bar q} \gamma_\lambda
(\beta_{0R} +\beta_{3R} \tau_3) (1+ \gamma_5) q
\nonumber \\
 & - & {\bar q} \gamma_\lambda
(\beta_{0L} +\beta_{3L} \tau_3) (1 - \gamma_5) q \} ({\bar \chi}_1
\gamma^\lambda
\gamma^5 \chi_1 \}
 \label{eq:eg 21}
\earr
with
\barr
\beta_{0R} &=& \Big( \frac {4} {9} \chi^2_{{\tilde u}_R}
+\frac {1} {9} \chi^2_{{\tilde d}_R}\Big) |C_{11} tan \theta_W|^2\nonumber \\
\beta_{3R} &=& \Big( \frac {4} {9} \chi^2_{{\tilde u}_R}
-\frac {1} {9} \chi^2_{{\tilde d}_R} \Big) |C_{11} tan \theta_W|^2
\label{eq:eg 22}\\
\beta_{0L} &=& | \frac {1} {6} C_{11} tan\theta_W
+\frac{1}{2} C_{21}|^2 \chi^2_{{\tilde u}_L} + | \frac {1} {6} C_{11}
tan\theta_W  - \frac{1}{2} C_{21}|^2 \chi^2_{{\tilde d}_L} \nonumber \\
\beta_{3L} &=& | \frac {1} {6} C_{11} tan\theta_W +\frac{1}{2} C_{21}|^2
\chi^2_{{\tilde u}_L} - | \frac {1} {6} C_{11} tan\theta_W
-\frac{1}{2} C_{21}|^2 \chi^2_{{\tilde d}_L} \nonumber
\earr
with
\beq
\chi^2_{{\bar q}_L} =  \frac {m_W^2}{m_{{\bar q}^2_L}-q^2} , \,\,\,\,
\chi^2_{{\bar q}_R} =  \frac {m_W^2}{m_{{\bar q}^2_R}-q^2} , \,\,\,
{\tilde q} = {\tilde u},{\tilde d}
 \label{eq:eg 23}
\eeq
The above parameters are functions of the four-momentum transfer which
in our case is negligible.  Proceeding as above we can obtain the effective
Lagrangian at the nucleon level as
  \beq
{\it L}_{eff} = - \frac {G_F}{\sqrt 2} ({\bar \chi}_1 \gamma^{\lambda}
\gamma^5 \chi_1) J_{\lambda} ({\tilde q})
 \label{eq:eg 24}
\eeq
 \beq
J_{\lambda}({\tilde q}) = {\bar N} \gamma_{\lambda} \{f^0_V({\tilde q})
+ f^1_V
({\tilde q}) \tau_3 + f^0_A({\tilde q}) \gamma_5 + f^1_A({\tilde q})
\gamma_5
\tau_3) N
  \label{eq:eg 25}
\eeq
with
\beq
 f^0_V = 6(\beta_{0R}-\beta_{0L}) , \,\,\, f^1_V = 2
(\beta_{3R}-\beta_{3L}), \,\,\,  f^0_A = 2g_V (\beta_{0R}+\beta_{0L}), \, \,
 f^1_A = 2g_A(\beta_{3R}+\beta_{3L}) \label{eq:eg 25a}
\eeq
We should note that this interaction is more suppressed than the ordinary
weak interaction by the fact that the masses of the s-quarks are usually
larger than that of the gauge boson $Z^0$.
In the limit in which the LSP is a pure bino ($C_{11} = 1,  C_{21} = 0$) we
obtain
\barr
\beta_{0R} &=& tan^2 \theta_W \Big( \frac {4} {9} \chi^2_{u_R}
+\frac {1} {9} \chi^2_{{\tilde d}_R}\Big) \nonumber \\
\beta_{3R} &=& tan^2 \theta_W \Big( \frac {4} {9} \chi^2_{u_R}
-\frac {1} {9} \chi^2_{{\tilde d}_R} \Big) \label{eq:eg 24}\\
\beta_{0L} &=&  \frac {tan^2 \theta_W} {36} (\chi^2_{{\tilde u}_L} +
 \chi^2_{{\tilde d}_L}) \nonumber \\
\beta_{3L} &=&  \frac {tan^2 \theta_W} {36}
(\chi^2_{{\tilde u}_L} -  \chi^2_{{\tilde d}_L})\nonumber
\earr

Assuming further that $\chi_{{\tilde u}_R} = \chi_{{\tilde d}_R}
= \chi_{{\tilde u}_L} = \chi_{{\tilde d}_L}$ we obtain
\barr
 f^1_V ({\tilde q}) / f^0_V({\tilde q}) &\simeq& + \frac{2}{9}
\nonumber \\
  f^1_A ({\tilde q})/ f^0_A({\tilde q}) &\simeq& + \frac{6}{11}
 \label{eq:eg 25}
\earr
If, on the other hand, the LSP were the photino ($C_{11} = cos \theta_W,
C_{21} = sin \theta_W, C_{31} = C_{41} = 0$) and the s-quarks were
degenerate there would be no coherent contribution ($f^0_V = 0$ if
$\beta_{0L} =\beta_{0R}$).\\
\vspace{0.2cm}

{\bf 2.2.3. The intermediate Higgs contribution}

The process (3) can be mediated via the physical intermediate Higgs
particles (see fig. 2). The relevant interaction can arise out of the
Higgs-Higgsino-gaugino interaction which takes the form
\barr
 {\it L}_{H\chi\chi} &=& \frac {g}{\sqrt 2} \Big({\bar{\tilde W}}_R
 {\tilde H}_{2L} H^{0*}_2 - {\bar{\tilde W}}_R {\tilde H}_{1L} H^{0*}_1
\nonumber \\
   &-&tan \theta_w ({\bar{\tilde B}}_R
  {\tilde H}_{2L} H^{0*}_2 - {\bar{\tilde B}}_R {\tilde H}_{1L} H^{0*}_1)
 \Big) + H.C.
 \label{eq:eg 26}
\earr
Proceeding as above we can express ${\tilde W}$ an ${\tilde B}$ in terms
of the appropriate eigenstates and retain the LSP to obtain
\barr
 {\it L} &=& \frac {g}{\sqrt 2} \Big((C_{21} -tan \theta_w C_{11})
C_{41}{\bar \chi}_{1R} \chi_{1L} H^{o*}_2 \nonumber \\
   &-&(C_{21} -tan \theta_w C_{11})
C_{31}{\bar \chi}_{1R} \chi_{1L} H^{o*}_1 \Big) + H.C.
 \label{eq:eg 27}
\earr
We can now proceed further and express the fields $H^{0*}_1$ and
$H^{o*}_2$ in terms of the fields $\varphi_1 = \chi_1, \varphi_2 = \chi_2,
\varphi_3 =-i\chi_3$.  The fourth field $\varphi_4 = -i\chi_4$
 has been eaten up the gauge boson (Goldstone
boson). $\chi_1$ and $\chi_2$ are
eigenfields of the real parts and  $\chi_3$ and $\chi_4$ are those of
imaginary parts.  They are normally designated as $h,H, A$ and $G$.
The results are, of course, model dependent.  At the tree level we obtain the
mixings $\theta_r$ and $\theta_i$ between the real and imaginary parts as
follows
\beq
tan 2 \theta_r = tan 2\beta \frac {2m^2_0 +m^2_Z} {2m^2_0 - m^2_Z}
 \label{eq:eg 28}
\eeq
\beq
tan 2 \theta_i = tan 2\beta
 \label{eq:eg 29}
\eeq
The masses are
\beq
m^2_1 = m^2_h = \frac{1}{2} [(m^2_0 + \frac{1}{2} m^2_Z)
-\sqrt{(m^2_0 + \frac{1}{2} m^2_Z)^2 -2m^2_0 m^2_Z cos^2 2\beta}]
 \label{eq:eg 30}
\eeq
\beq
m^2_2 = m^2_H = \frac{1}{2} [(m^2_0 + \frac{1}{2} m^2_Z)
+\sqrt{(m^2_0 + \frac{1}{2} m^2_Z)^2 -2m^2_0 m^2_Z cos^2 2\beta}]
 \label{eq:eg 31}
\eeq
\beq
m^2_3 = m^2_A = m^2_0 \qquad  with \,\,\, m^2_0 = - \mu B/sin 2\beta
 \label{eq:eg 32}
\eeq
We thus can write
\beq
H^{0*}_1 = \sum^3_{j=1} \xi^{(3)}_j \varphi_j , \,\,\,
H^{0*}_2 = \sum_{j=1} \xi^{(4)}_j \varphi_j
 \label{eq:eg 33}
\eeq
Even though one can now include radiative corrections [31] in our work we
found it adequate to use the above expressions and take $m_0$ and $tan
\beta$ from ref.[1].  The results are presented for the reader's
convenience in table III.

We thus obtain
\beq
{\it L} = \frac {g}{\sqrt 2} (C_{11} tan \theta_w -C_{21})
 \sum^3_{j=1}(C_{31} \xi^{(3)}_j - C_{41} \xi^{(4)}_j)
 \label{eq:eg 34}
\eeq
For the quark vertex we need the Yukawa interactions
\beq
{\it L}_Y = f^u_{ij} {\bar u}^0_{iL} u^0_{jR} H^{0*}_2
+f^d_{ij} {\bar d}^0_{iL} d^0_{jR} H^{0*}_1 + H.C.
 \label{eq:eg 35}
\eeq
In terms of the physical fields we obtain
\barr
{\it L}_Y &=& mf^u_i {\bar u}_{iL} u_{iR} + \frac {g}{2\sqrt 2}
\frac {m^u_i}{m_W sin\beta}  {\bar u}_{iL} u_{iR} \varphi_j\nonumber \\
&+& m^d_i {\bar d}_{iL} d_{iR} +  \frac {g}{2\sqrt 2}
\frac {m^i_d}{m_W sin\beta}  {\bar d}_{iL} d_{iR} \varphi_j
 \label{eq:eg 36}
\earr
(summation over $i$ and $j$ is understood). Combining the above results
we obtain
the four-fermion interaction \beq
{\it L}_{eff} = - \frac {G_F}{\sqrt 2}
 {\bar \chi}_1 \chi_1 {\bar q} [f^+_s + f^-_s \tau_3] q
 \label{eq:eg 37}
\eeq
with
\beq
f^\pm_s = 2((C_{11} tan \theta_w -C_{21})
 \sum^3_{j=1}(C_{41} \xi^{(4)}_j - C_{31} \xi^{(3)}_j)
( \frac {m_u}{m_w sin\beta} \xi^{(4)}_j \pm  \frac {m_d}{m_w cos\beta}
 \xi^{(3)}_j)
 \label{eq:eg 38}
\eeq
In order to reduce this to the nucleon level we follow the work of Addler
[30] which leads to
\beq
{\it L}_{eff} = - \frac {G_F}{\sqrt 2}
 {\bar \chi}_1 \chi_1 {\bar N} [f^0_s + f^1_s \tau_3] N
 \label{eq:eg 39}
\eeq
with
\beq
 f^0_s = 1.86 f^+_s , \,\, f^1_s = 0.48 f^-_s
 \label{eq:eg 40}
\eeq
 \vspace{0.2cm}

{ \it {\bf 3. Evaluation of the nuclear matrix elements}}.

Combining for results of the previous section we can write
 \beq
{\it L}_{eff} = - \frac {G_F}{\sqrt 2} \{({\bar \chi}_1 \gamma^{\lambda}
\gamma_5 \chi_1) J_{\lambda} + ({\bar \chi}_1
 \chi_1) J\}
 \label{eq:eg 41}
\eeq
where
\beq
  J_{\lambda} =  {\bar N} \gamma_{\lambda} (f^0_V +f^1_V \tau_3
+ f^0_A\gamma_5 + f^1_A\gamma_5 \tau_3)N
 \label{eq:eg 42}
\eeq
with
\barr
 f^0_V  &=&  f^0_V(Z) + f^0_V({\tilde q}),  \qquad
f^1_V  =  f^1_V(Z) + f^1_V({\tilde q})
\nonumber \\
  f^0_A  &=&  f^0_A(Z) + f^0_A({\tilde q}) , \qquad
f^1_A  =  f^1_A(Z) + f^1_A({\tilde q})
 \label{eq:eg 43}
\earr
and
\beq
  J =  {\bar N} (f^0_s +f^1_s \tau_3) N
\nonumber \\
\eeq
By performing a straightforward trace calculation we obtain
\barr
|{\it m}|^2 &=& \frac{1}{m_1^2} \{ (p^\lambda_f J_\lambda)(p^\mu_i J^*_\mu)
+ (p^\lambda_i J^\mu_\lambda)(p^\mu_f J^*_\mu)
-J^\lambda J^*_\lambda p^\mu_i (p_f)_\mu \nonumber \\
 &-& m_1^2 J^\lambda J^*_\lambda + p_ip_f|J|^2 \}
\label{eq:eg 44}
 \earr
By noting that the LSP is  an extremely non relativistic particle ($\beta
\leq 10^{-3}$) we retain the leading term for each type of matrix element
to get
\barr
|{\it m}|^2 &=& \frac{1}{m_1^2} \{\frac{1}{2} (E_f E_i -m_1^2 +
{\bf p}_i\cdot {\bf p}_f) |J_0|^2 + \frac{1}{2} (E_i E_f +m_1^2) |{\bf J}|^2
+ E_i E_f |J|^2 \} \nonumber \\
& \simeq & \frac{(E_f E_i -m_1^2 + {\bf p}_i\cdot {\bf p}_f)}{m^2_1} J^2_0
+ |{\bf J}|^2 + |J|^2
\label{eq:eg 45}
 \earr
where $E_i, {\bf p}_i, E_f, {\bf p}_f$ and $m_1$ are the
kinematical variables of LSP (in the laboratory frame).

The first and the last matrix elements $J_0$ and $J$ are non zero even for
$0^+ \ra 0^+$ transitions. Furthermore, all nucleons participate
coherently and we expect the matrix elements of  $J_0$ and $J$ to be
proportional to the mass number A.  The matrix element of J is expected
to be smaller than that of $J_0$ due to the smallness of the quark masses
(see previous section).  The coefficient of $J_0$, however, is quite
small for non relativistic LSP's (the opposite sign of $m^2_1$ is a
consequence of the Majorana nature of LSP).  It is proportional to
$\beta=\upsilon/c$.  Finally the matrix element of ${\bf J}$ vanishes for
$0^+ \ra 0^+$ transitions (to leading order).  Even for $J^\pi \neq 0$
it is expected to be smaller than that of $J_0$ especially for heavy nuclei,
 since
not all nucleons participate (non coherence).

{}From the above discussion we conclude that, due to the Majorana nature of
LSP, the matrix element $|{\it m}|^2$ is suppressed.  Therefore, all three
matrix elements need be considered.  One can easily find
\beq
 |J_0|^2 = A^2 |F({\bf q}^2)| ^2 [f^0_V -f^1_V \frac{N-Z}{A}]^2
\label{eq:eg 46}
\eeq
\beq
 J^2 = A^2 |F({\bf q}^2)| ^2 [f^0_s -f^V_s \frac{N-Z}{A}]^2
\label{eq:eg 47}
\eeq
where $F({\bf q} ^2) \simeq 1$ is the nucleon form factor for momentum
transfer  ${\bf q}$. Also
\beq
 |{\bf J}|^2 = \frac{1}{2J_i+1} |<J_i ||{\vec \sigma}||J_i>|^2 [g^0_A
\pm g^1_A]^2 \label{eq:eg 48}
\eeq
where the $+(-)$ sign is associated for protons or neutron holes (neutrons
or proton holes).\\
The nuclear matrix element $<J_i ||{\bf \sigma}||J_i>$
vanishes for $J_i =0^+$. For $J_i \neq 0^+$ it depends on the details of the
structure of the nucleus.  For $^{207}Pb$, however, it can easily be
evaluated, since it is a single particle configuration (one $2p_{1/2}$
neutron hole outside the closed shell).
 For a single particle configuration we get

\beq
 \frac{1}{2j+1} |<\ell j ||{\vec \sigma}||\ell j>|^2  =
\cases{ j/(j+1),   \cr  (j+1)/j,  \cr} \quad
 {j=\ell-1/2    \atop   j=\ell+1/2 }
\label{eq:eg 49}
\eeq

Thus for $^{207}Pb$ we obtain
\beq
 |{\bf J}| = \frac{1}{3} (f^0_A + f^1_A)^2 \label{eq:eg 50}
\eeq
For the other odd nuclei which are  relevant as targets in searching for dark
matter the situation is not so simple.  Detailed calculations are under
way.
For the time being we will present estimates for the three light nuclei
($^3_2He$,
$^{19}_9F$ and $^{23}_{11}Na$).  We will assume [32] that the space wave
function has symmetry characterized  by a Young
Tableaux $[f]$ which is  as symmetric as possible.  This is due to the fact
that the two nucleon interaction is attractive and short -ranged.  It thus
favors nucleon pairs in which the nucleons are as close as possible.  The
spin-isospin wave function is characterized by symmetry $[{\tilde f}]$
($[{\tilde f}]$ is obtained from $[f]$ by interchanging rows and columns [32]).
This guarantees that the total wave function is  antisymmetric.  The
orbital angular monentum is assumed to be the lowest allowed.

1.  The $^3_2He$ target.  The wave function is assumed to be spatially
symmetric, $[f] =[3]$, i.e. of the form \beq
 \psi(gs) = [3] L=0 ; [1^3] s=\frac{1}{2},\quad I = \frac{1}{2},\quad I_3 = -
\frac{1}{2}
 ; J = \frac{1}{2}
 \nonumber
\eeq
The isoscalar matrix element vanishes while the isovector matrix element is
\beq
|{\bf J}|^2 = 27 |f^1_A|^2 \nonumber
\eeq

2.The $^{19}_9F$ target.  The wave function is described in terms of three
nucleons outside the closed shell $^{16}_8O$ nucleus, i.e.
\beq
 \psi(gs) = [3](60) L=0 ; [1^3] s=\frac{1}{2},\quad  I = \frac{1}{2} ,\quad I_3
=-\frac{1}{2}
 ; J = \frac{1}{2}
 \nonumber
\eeq
where SU(3) representation (60) has the largest value of the Casimir
invariant [33].  The obtained matrix element is the same as above, i.e.
 \beq
|{\bf J}|^2 = 27 |f^1_A|^2 \nonumber
\eeq

3. The  $^{23}_{11}Na$ system. It is described as 7 particles outside the
closed
shell.  The spatial symmetry is assumed to be $[f] = [43]$, i.e.
 \beq
 \psi(gs) = [43](83) L=1 ; [2221] s=\frac{1}{2},\,  I = \frac{1}{2} ,\, I_3
=-\frac{1}{2}; J = \frac{3}{2} \nonumber
\eeq
The SU(4) spin - isospin symmetry is equivalent to $[1^3]$.  The  matrix
element is suppressed by the factor dim [4,2]/ dim [4,3], [42] and [4,3]
viewed as representation of the symmetric groups [32] $S_6$ and $S_7$.  This
ratio is 9/14.  Furthermore, an angular momentum reduction coefficient of 1/9
enters yielding
 \beq
|{\bf J}|^2 = \frac {9}{14} \frac {1}{9} 27 |f^1_A|^2 = \frac {27}{14}
 |f^1_A|^2
 \nonumber
\eeq

At this point we should mention that for the extra
non-relativistic process (3) traditional nuclear physics techniques
should be more
reliable than attempts to extract $|{\bf J}|^2$ [15] from the EMC data [34].

\vspace{0.2cm}
{}
{ \it {\bf 4. Cross-Sections}}.

With the above ingredients the differential cross section can be easily
calculated. For the benefit of the experimentalists we prefer to present our
results in the laboratory frame.  The scattering is in the forward direction
 $\xi =
{\hat p}_i \cdot {\hat q} \geq 0, ({\bf p}_i$ is the initial LSP momentum,
${\bf q}$ the momentum transferred to the nucleus).
After making the non-relativistic  approximation one finds that
\barr
\frac{d\sigma}{d \Omega} &=& \frac{\sigma_0}{\pi} (\frac{m_1}{m_p})^2
\frac{1}{(1+\eta)^2} \xi \theta (\xi) \{ A^2 [\beta^2(f^0_V - f^0_A
\frac{N-Z}{A})^2 \times \nonumber \\
& \times & (1 - \frac{2\eta+1}{(1+\eta)^2} \xi^2) + (f^0_s - f^1_s
\frac{N-Z}{A})^2]
 \nonumber \\
&+& \frac{1}{2J_i+1} <J_i ||{\vec \sigma}(f^0_A+\tau_3f^1_A)||J_i>^2 \}
\label{eq:eg 51}
 \earr
where $m_p$ is the proton mass, $\eta = m_1/m_A$ ($m_A$ is the mass of
the nucleus), $\beta = \upsilon/c$ ($\upsilon$ is the velocity of LSP) and
\beq
\sigma_0 = \frac{1}{2\pi} (G_F m_p)^2 = 0.77 \times 10^{-38}
cm^2 \label{eq:eg 52}
\eeq
The total cross-section becomes
\barr
\sigma &=& \sigma_0 (\frac{m_1}{m_p})^2 \frac{1}{(1+\eta)^2}
 \{ A^2 [\beta^2 (f^0_V - f^1_V \frac{N-Z}{A})^2
\frac{2\eta^2+2\eta +1}{2(1+\eta)^2}
  \nonumber \\
&+&  (f^0_s - f^1_s \frac{N-Z}{A})^2] +
\frac{1}{2J_i+1} |<J_i||(f^0_A+\tau_3 f^1_A)\sigma||J_i>^2 \}
\label{eq:eg 53}
 \earr

We notice that all exchanges (Z, s - quark and  Higgs) allow for a
coherent contribution of all nucleons yielding a matrix element
proportional to the
nuclear mass A.  The Z and s-quark contribution is suppressed by a factor
$\beta^2$ for  a non-relativistic LSP which, as we have already mentioned,
 is due
to the majorana nature of LSP [11,12].  The parameters $f^0_V,
 f^0_A$ and $f^1_A$
as well as the ratios of the isovector to the isoscalar coefficients,
 $f^1_V/f^0_V$
and $f^1_s|f^0_s$, are presented in table IV.

{}From the data of this table we can draw the following general conclusions:

1. For $0^+ \ra 0^+$ transitions the Higgs contribution becomes  dominant for
solutions 1 and 2 in spite of the smallness of the quark masses ($m_u=5 MeV,
m_d = 10MeV$).  For solution 3 the Higgs contribution becomes comparable to
the combined effect of the Z and s-quark exchange.

2. The isovector contribution is small in all cases and additive to the
isoscalar contribution.  The
isovector contribution of the Z-exchange is partly  cancelled by that of the
s-quark contribution.

3.  For odd nuclear targets ($J_i \neq 0$)  the spin contribution becomes
important.  This contribution, which
arises from the Z and s-quark exchanges, is not suppressed by the majorana
nature of the LSP. It will dominate unless the matrix element of the axial
current is accidentally suppressed.  Since this does not scale with $A^2$,
it is somewhat less
important in the case of heavy nuclear targets.

4.  The coherent Z and s-quark contribution has an extra angular dependence,
which, given a sufficient number of events, could be used for its experimental
discrimination.\\

\vspace{0.2cm}
{}
{ \it {\bf 5. Discussion of the results}}.

Using the formulas given above and the data of tables I-IV we can compute the
total cross - sections for the LSP scattering with nuclei which can be used as
targets. We see that the cross - section for  $0^+ \ra 0^+$ transitions,
as well as
and the coherent part of the cross - section for odd-mass nuclei,
tend to inrcrease
with square of the mass number A.  The spin matrix element does
not show such an
increase and depends on the details of the nuclear structure.  It has been
reliably calculated only in the case of $^{207}Pb$ which is a
 single nucleon hole
outside the closed shell.  Our numerical results are shown in tables Va
(for Z and s-quark exchange) and Vb (for Higgs exchange).  From these tables,
 we verify that for the coherent part the Higgs contribution, even though the
quarks are very light, becomes dominant for solutions 1 and 2).
For solution 3
the Z and s-quark exchange competes with the Higgs contribution.  The largest
cross - section is obtained for $^{207}Pb$.  Indeed for $^{207}Pb$ we get\\
Solution 1 : $8.1 \times 10^{-38} cm^2$ \\
Solution 2 : $1.9 \times 10^{-37} cm^2$ \\
Solution 3 : $2.6 \times 10^{-38} cm^2$\\
Unfortunately this cross section is extremely small.  This makes the detection
of LSP extremely difficult.  Indeed the event rate is given by [16]
\beq
 \frac {dN}{dt} = \frac{\rho_x}{m_1} <\upsilon> \sigma \frac{m}{Am_p}
\nonumber
\eeq
where $\rho_x$ is the density of LSM in our solar system, $m,A$ are the mass
and the mass number of the target, $m_1$ is the mass of LSP and $<\upsilon>$
its average velocity.  We find
\beq
 \frac {dN}{dt} = 5.0 day^{-1} \frac{\rho_x}{0.3 GeV/cm^3} \frac{100}{m_1GeV}
\frac{m}{1Kg} \frac{V}{320 Km/s}
 \frac{\sigma}{10^{-34}cm^2}\nonumber
\eeq

Using $\rho_x=0.3 GeV/cm^3, \qquad m = 1 Kg$ and
 $V=320 Km$  [15,16],and our results $m_1 = 27 GeV$ and $\sigma = 1.9
\times 10^{-37}
cm^2$ (solution 2) for $^{207}_{82}Pb$  we obtain
\beq
 \frac {dN}{dt} = 6.6 \times 10^{-3} events|day \simeq
2.5 events /year \nonumber
\eeq

Finally we should remark that, even though, as we have mentioned
earlier, the predictions of SUSY have become quite constrained and reliable in
recent  years, the calculated cross - section for process (3) has some
uncertainties in it.  It depends on the inverse fourth power of
the s-quarks and
Higgs particles.  It also depends on the small mixings $C_{41}$ and $C_{31}$
(eqs.
(14)-(17) for Z-exchange and eq. (43) for Higgs exchange).  In spite of this,
within the allowed parameter space with rather wide variations, e.g.
 $tan \beta$
ranging from 1.5 to 10,  the variation in the calculated cross section
 is not very
large.  The nuclear matrix elements for the coherent process,
 which is all there
is for $0^+$ targets, is very accurate.  The evaluation of the spin matrix
element
for odd targets is less accurate but it can be reliably estimated, at
least for
$^{207}Pb$.  So our estimate for the cross section should be
 viewed as quite
reliable.  Thus, barring completely unforseen circumstances,
the event rate is
expected to be small.


\newpage


\newpage

{\bf Figure Captions}

Fig. 1.  Two diagrams which contribute to the elastic scattering of LSP with
Nuclei. Z-exchange (fig. 1a) and s-quark exchange (fig. 1b).  Due to the
Majorana
nature of LSP only its pseudovector coupling contributes. $J_\lambda$ can be
parametrized in terms of  four form factors $f^0_V, f^1_V, f^0_A, f^1_A$.

Fig. 2.  The same as in fig. 1, except that the intermediate Higgs exchange is
considered. This leads to an effective scalar interaction with two form factors
$f^0_S$, (isoscalar) and  $f^1_S$ (isovector)

\newpage

\begin{table}
{\bf Table I} : SUSY parameters which are relevant for the scattering of
LSP with nuclei.  They were taken from ref. [1].\

\begin{tabular}{|r|l|c|c|c|c|c|c|c|c|c|}
\hline
Solution &   $tan \beta$    &  $ \mu$   &  $M_1$    &   $M_2$  &
$m_{{\tilde u}_R}$   &   $m_{{\tilde d}_R}$ & $m_{{\tilde u}_L}$  & $m_{{\tilde
d}_L}$ & $m_1$ \\
& & (GeV) & (GeV) & (GeV) & (GeV) & (GeV) & (GeV) & (GeV) & (GeV)\\
\hline
     1     &    10   & 450 &  126   &  245 & 677  &  676 &700&705&126 \\
     2     &    1.5  & -218 &  45   &  90 & 276   &  276 &283&288&27 \\
     3     &    5    & 304 &  102   &  200 & 551  &  550 &570&575&102 \\
\hline
\end{tabular}
\end{table}


\addvspace{2.5in}


\begin{table}
{\bf Table II} : The relevant components $C_{j1}, j =
1,2,3,4$ of LSP or $\chi_1$  (see eq. (10)) and its
masses $m_1$ obtained from the data of table I.\

\begin{tabular}{|l|c|c|c|c|c|}
\hline
& Solution 1 &  Solution 2 & Solution 3 \\
\hline
     $C_{11}$ &   .9945 & .8225 &  .9891 \\
     $C_{21}$ &   -.5779$\times 10^{-2}$ & -.4343 &-.6258$\times 10^{-2}$ \\
    $C_{31}$ &   .1029 & .2968 &  -.1458 \\
 $C_{41}$ &   -.1897$\times 10^{-1}$ & -.2164 &-.2136$\times 10^{-1}$\\
\hline
\end{tabular}
\end{table}

\noindent

\begin{table}
{\bf Table III} : The mass of the physical Higgs particles $\varphi_i$ and the
coefficients $\xi^{(3)}_j$ and $\xi^{(4)}_j$ in the decomposition of the
neutral Higgs particles $H^{0*}_1$ and $H^{0*}_2$ i.e. $H^{0*}_1 =
 \sum_j \xi^{(3)}_j \varphi_j$ and $H^{0*}_2 =
 \sum_j \xi^{(4)}_j \varphi_j$ (j = 1,2 correspond to the real parts and j = 3
to the imaginary part).

\begin{tabular}{|l|c|c|c|c|c|}
\hline
Variable& Solution 1 &  Solution 2 & Solution 3 \\
\hline
     $m_1(GeV)$ &   68.7 & 34.5 &  85.7 \\
     $m_2(GeV)$ & 119 & 221 & 197 \\
     $m_3(GeV)$ & 130 & 217 & 201 \\
\hline
     $\xi^{(3)}_1$ &  0.5149 & 0.6100 &  0.4581 \\
     $\xi^{(3)}_2$ &  -0.4847 & -0.3877 & -0.5385 \\
     $\xi^{(3)}_3$ & 0.7036 & 0.5883 & 0.6934 \\
\hline
     $\xi^{(4)}_1$ &  0.4847 & 0.3877 &  0.5385 \\
     $\xi^{(4)}_2$ &  0.5149 & 0.6100 & 0.4581 \\
     $\xi^{(4)}_3$ & 0.0704 & 0.3922 & 0.0139 \\
\hline
\end{tabular}
\end{table}

\noindent

\begin{table}
{\bf Table IV} : The parameters  $\beta f^0_V, f^0_S, f^0_A, f^1_A$ and
$f^1_V/f^0_V, f^1_S/ f^0_S$ for three SUSY solutions (see text).  The value
of $\beta = 10^{-3}$ was used.  Also in the evaluation of $f^0_S$ and $f^1_S$
we
used $m_u = 5 MeV$ and $m_d = 10 MeV$ for the quark masses.  \

 \begin{tabular}{|l|c|c|c|c|c|}
\hline
quantity& Solution 1 &  Solution 2 & Solution 3 \\
\hline
     $\beta f^0_V(Z)$ & $0.475\times10^{-5}$ & $1.916\times10^{-5}$
& $0.966\times10^{-5}$ \\
     $f^1_V(Z)/f^0_V(Z)$ & -1.153 & -1.153 & -1.153\\
     $\beta f^0_V({\tilde q})$ &$1.271\times10^{-5}$&$0.798\times10^{-5}$
 & $1.898\times10^{-5}$ \\
$f^1_V({\tilde q})/f^0_V({\tilde q})$ & 0.222 & 2.727 &0.217 \\
$\beta f^0_V$ &$1.746\times10^{-5}$&$2.617\times10^{-5}$
 & $2.864\times10^{-5}$ \\
$f^0_V/f^1_V$ &-0.153 &-0.113 &-0.251 \\
$f^0_S$ &$1.71\times10^{-5}$&$8.02\times10^{-4}$ &$-5.51\times10^{-5}$ \\
$f^1_S/f^0_S$ &-0.24 &-0.15 &-0.25 \\
$f^0_A(Z) $ &- &- &- \\
$f^1_A(Z) $ & $1.27\times10^{-2}$&$5.17\times10^{-2}$
 & $2.58\times10^{-2}$ \\
$f^0_A({\tilde q}) $ & $0.510\times10^{-2}$&$3.55\times10^{-2}$
 & $.704\times10^{-2}$ \\
$f^1_A({\tilde q}) $& $0.277\times10^{-2}$&$0.144\times10^{-2}$
 & $0.423\times10^{-2}$ \\
$f^0_A $& $0.510\times10^{-2}$&$3.55\times10^{-2}$
 & $.704\times10^{-2}$ \\
$f^1_A $& $1.55\times10^{-2}$&$5.31\times10^{-2}$
 & $3.00\times10^{-2}$ \\
\hline
\end{tabular}
\end{table}

\noindent

\begin{table}
{\bf Table Va} : The isospin correction  $IC =| 1 - \frac {f^1_V}{f^0_V}
\frac {N-Z}{A}|^2$ and the total coherent correction associated with Z-boson
 and
s-quark exchange for various nuclei of interest.  For some odd mass nuclei we
also present in parenthesis the cross - section associated with the spin matrix
elements $|{\bf J}|^2$.\

 \begin{tabular}{|l|c|c|c|c|c|c|c|c|}
\hline
Nucleus&\multicolumn {2}{|c|} { Solution 1} & \multicolumn {2}{|c|}{Solution
2} & \multicolumn {2}{|c|} {Solution 3} \\

&IC&$\sigma (cm^2)$&IC&$\sigma (cm^2)$&IC&$\sigma (cm^2)$\\
\hline
 $^3_2He$ &0.90& $1.8\times10^{-46}$ &0.93& $3.1\times10^{-46}$&0.84
& $4.6\times10^{-46}$ \\
& & $(1.5\times10^{-38})$ & & $(7.9\times10^{-38})$&
& $(5.4\times10^{-38})$ \\
\hline
 $^{19}_9Fe$ &1.02& $2.3\times10^{-43}$ &1.01& $1.8\times10^{-43}$&1.03
& $5.8\times10^{-43}$ \\
& & $(1.5\times10^{-38})$ & & $(7.9\times10^{-38})$&
& $(5.4\times10^{-38})$ \\
\hline
 $^{23}_{11}Na$ &1.01& $4.6\times10^{-43}$ &1.01& $3.2\times10^{-43}$&1.02
& $1.1\times10^{-42}$ \\
& & $(1.1\times10^{-39})$ & & $(5.6\times10^{-39})$&
& $(3.9\times10^{-39})$ \\
\hline
 $^{40}_{20}Ca$ &1.00& $3.1\times10^{-42}$ &1.00& $1.5\times10^{-42}$&1.00
& $7.0\times10^{-42}$ \\
\hline
 $^{71}_{31}Ga$ &1.04& $2.0\times10^{-41}$ &1.03& $6.4\times10^{-42}$&1.07
& $4.5\times10^{-41}$ \\
\hline
 $^{72}_{32}Ge$ &1.03& $2.1\times10^{-41}$ &1.03& $6.4\times10^{-42}$&1.06
& $4.6\times10^{-41}$ \\
\hline
 $^{75}_{33}As$ &1.04& $2.4\times10^{-41}$ &1.03& $7.0\times10^{-42}$&1.06
& $5.2\times10^{-41}$ \\
\hline
 $^{76}_{33}Ge$ &1.05& $2.5\times10^{-41}$ &1.04& $7.5\times10^{-42}$&1.08
& $5.7\times10^{-41}$ \\
\hline
 $^{127}_{53}I$ &1.05& $1.1\times10^{-40}$ &1.04& $2.5\times10^{-41}$&1.08
& $2.4\times10^{-40}$ \\
\hline
 $^{134}_{54}Xe$ &1.06& $1.4\times10^{-40}$ &1.04& $2.8\times10^{-40}$&1.09
& $2.8\times10^{-40}$ \\
\hline
 $^{207}_{82}Pb$ &1.07& $4.2\times10^{-40}$ &1.05& $7.6\times10^{-41}$&1.11
& $8.9\times10^{-40}$ \\
& & $(7.6\times10^{-39})$ & & $(1.3\times10^{-38})$&
& $(1.9\times10^{-38})$ \\
\hline
\end{tabular}
\end{table}

\noindent

\begin{table}
{\bf Table Vb} : The isospin correction  $IC =| 1 - \frac {f^1_S}{f^0_S}
\frac {N-Z}{A}|^2$ and the total cross - section associated with Higgs
exchange in the LSP scattering with nuclei.\

 \begin{tabular}{|r|l|c|c|c|c|c|c|c|}
\hline
Nucleus&\multicolumn {2}{|c|} { Solution 1} & \multicolumn {2}{|c|}{Solution
2} & \multicolumn {2}{|c|} {Solution 3} \\

&IC&$\sigma (cm^2)$&IC&$\sigma (cm^2)$&IC&$\sigma (cm^2)$\\
\hline
 $^3_2He$ &0.846& $1.9\times10^{-44}$ &0.903& $3.4\times10^{-43}$&0.840
& $1.8\times10^{-45}$ \\
 $^{19}_9Fe$ &1.03& $2.6\times10^{-41}$ &1.02& $2.6\times10^{-40}$&1.03
& $2.5\times10^{-42}$ \\
 $^{23}_{11}Na$ &1.02& $3.3\times10^{-41}$ &1.01& $4.8\times10^{-40}$&1.02
& $3.4\times10^{-42}$ \\
 $^{40}_{20}Ca$ &1.00& $3.5\times10^{-40}$ &1.00& $2.4\times10^{-39}$&1.00
& $3.4\times10^{-41}$ \\
 $^{71}_{31}Ga$ &1.06& $2.9\times10^{-39}$ &1.03& $1.2\times10^{-38}$&1.06
& $2.6\times10^{-40}$ \\
 $^{72}_{32}Ge$ &1.05& $3.0\times10^{-39}$ &1.04& $1.2\times10^{-38}$&1.05
& $2.6\times10^{-40}$ \\
 $^{75}_{33}As$ &1.06& $3.2\times10^{-39}$ &1.03& $1.3\times10^{-38}$&1.06
& $2.8\times10^{-40}$ \\
 $^{76}_{33}Ge$ &1.08& $3.2\times10^{-39}$ &1.05& $1.3\times10^{-38}$&1.08
& $2.9\times10^{-40}$ \\
 $^{127}_{53}I$ &1.08& $1.8\times10^{-38}$ &1.05& $4.7\times10^{-38}$&1.08
& $1.5\times10^{-39}$ \\
 $^{134}_{54}Xe$ &1.10& $1.9\times10^{-38}$ &1.06& $5.3\times10^{-38}$&1.10
& $1.7\times10^{-39}$ \\
 $^{207}_{82}Pb$ &1.10& $7.3\times10^{-38}$ &1.06& $1.8\times10^{-37}$&1.10
& $5.8\times10^{-39}$ \\
\hline
\end{tabular}

\end{table}



\begin{thebibliography}{99}
\bibitem{KK}G.L. Kane, Chris Kolda, Leszek Roszkowski and James D. Wells,
Phys. Rev. D{\bf 49} (1994)6173
\bibitem{BB}V. Barger, M.S. Berger and P. Ohmann, Phys. Rev.
D{\bf 49} (1994) 4908
\bibitem{SM}Stephen P. Martin and Pierre Ramond, Sparticle Spectrum
Constraints, NUB-3067-93TH, UFIFT-HEP-9316, SSCL-Preprint- 439.

D. Castano, E.J. Piard and P. Ramond, Phys. Rev. {\bf 49} (1994) 4882
 \bibitem{N}H.P. Niles, Phys. Rep. {\bf 110} (1984) 1
\bibitem{HK}H.E. Haber and G.L. Kane, Phys. Rep. {\bf 117} (1985) 75
\bibitem{LN}A. B. Lahanas and D. V. Nanopoulos,  Phys. Rep. {\bf 145} (1987)1
\bibitem{NA}P. Nath, R. Arnowitt  and A.H. Chansedine, Applied N=1
Supersymmetry (World Scientific, Singapore (1984))
\bibitem{LNZ}J.Lopez, D.V. Nanopoulos and A. Zichichi, Prog.
Part.  Nucl. Phys.{\bf 33} (1994) 303
\bibitem{B} W. de Boer, ibid p201
 \bibitem{KLV92}T.S. Kosmas, G.K. Leontaris and J.D. Vergados,Prog.
Nuc. Part.  Phys. {\bf 33} (1994) 397
 \bibitem{GW}W. Goodman and E. Witten, Phys. Rev.
 D {\bf 31} (1985)3059
\bibitem{G}K. Griest, Phys. Rev. Lett. {\bf 62} (1988) 666 \\
	Phys. Rev. D {\bf 38} (1988)2357; D {\bf 39} (1989) 3802
\bibitem{BFG}R. Barbieri, M. Frigini and G.F. Giudice,
Nucl. Phys. {\bf B313} (1989) 725
\bibitem{EF} J. Ellis, and R. Flores, Phys. Lett. B {\bf 263} (1991) 259,
B {\bf 300} (1993) 175; Nucl. Phys. {\bf B400} (1993) 25
\bibitem{BKK}V.A. Bednyakov, H.V. Klapdor-Kleingrothaus and S.G. Kovalenko,
 Phys. Lett.  {\bf B329} (1994) 5
\bibitem{AN}Searching for SUSY Dark Matter, R. Arnowitt and P. Nath,
CTP TAMU - 53/94, NUB-TH-3104/94, hep-ph 1904/1350

R. Arnowitt and Pram Nath, Sensitivity of Dark Matter Detectors to SUSY Dark
Matter, CERN-TH 7289/94
 \bibitem{SL}P.F. Smith and J. P. Lewin, Phys. Rep. {\bf
187} (1990) 203
\bibitem{G.S}G. Steigman, An Introduction to Cosmological Dark
Matter, Proc. Int. School on Cosm. Dark Matter, Valencia, Spain, 1993 p.1.
(ed.
J.W.F. Valle and A. Perez)
\bibitem{Ef}Evidence for Dark Matter, Michael Roman-Robinson, ibid p.7.
\bibitem{CF}The large Scale Structure of the Universe, Carlos S. Frenk, ibid
p.65.

Structure Formation in CDM an CHDM Cosmologies, Joel R. Primack, ibid p. 81

J.R. Primack et al., Ann. Rev. Nucl. Part. Sci. {\bf 38} (1988)751
\bibitem{Ma}Mixed Dark Matter and Low-energy Supersymmetry, A.
Masiero, ibid p.247.
 \bibitem{SIM}Looking for Distinctive Signals of Particle Dark Metter
A. Morales, E. Garcia, J. Morales, M.L. Sarsa, A. Ortiz de Solorzano, J.
Puimedon, C. Saenz, A-Solinas, J.A. Villar, S. Cerezo, F.T. Avignone, J.I.
Collar, R. L. Brodzinski, H.S. Miley, J.H. Reeves, Ibid p.259
\bibitem{COBE}COBE data, G. Smoot et al., Astrophys. J. {\bf 396} (1992)
L1
\bibitem{DC}Present and Future Underground Experiments, David B. Cline (and
references therein), ref. [18]
\bibitem{Mi}Detectors for Dark Matter Interactions Opretated at Low
Temperatures, Faus von Feilitzch, Int. Workshop on Neutrino Telescopes,
Valenzia Feb. 13-15, 1990 (ed. Milla Baldo Ceolin) p.257.
\bibitem{Ca}J.I. Callar et al., Phys. Rev.{\bf D47} (1993)5238
\bibitem{Dr}A. K. Druckier et al., Phys. Rev.{\bf D 33} (1986)3495

J.I. Callar et al., Phys. Lett.{\bf D 275} (1992)181
\bibitem{GV}G.J. Gounaris and J.D. Vergados, Phys. Rev. {\bf C17} (1978)1155
\bibitem{Ve} S.L. Adler et. al., Phys. Rev. {\bf D 11} (1975) 3309

J.D. Vergados, Phys. Lett.{\bf B184} (1987)55
\bibitem{Ok}L.B. Okun, Quarks and Leptons, section 29.3.4
\bibitem{Mi}Micheal S. Berger,  Phys. Rev. {\bf D41} (1990) 225
\bibitem{MH}Morton Hamermesh, Group Theory, Addison Wesley, Reading, Mass(1964)
\bibitem{JPE}J.P. Elliott, Prog.Roy. Soc. A245 (1958) 128, 562

K.T. Hecht, Nucl. Phys. {\bf 62} (1965) 1
\bibitem{JA}J. Ashman et al., EMC collaboration, Nucl. Phys. {\bf B328}
(1989) 1
\end{thebibliography}
 \end{document}